\documentclass[12pt]{article}

\usepackage{times}

\usepackage{url}

\usepackage{graphicx}

\newcommand{\captionfonts}{\small}
\makeatletter
\long\def\@makecaption#1#2{%
  \vskip\abovecaptionskip
  \sbox\@tempboxa{{\captionfonts #1: #2}}%
  \ifdim \wd\@tempboxa >\hsize
    {\captionfonts #1: #2\par}
  \else
    \hbox to\hsize{\hfil\box\@tempboxa\hfil}%
  \fi
  \vskip\belowcaptionskip}
\makeatother

\newenvironment{sciabstract}{%
\begin{quote} \bf}
{\end{quote}}

\newcounter{lastnote}

\newcommand{\defn}{\textbf}

\newcommand{\etal}{{\it{}et~al.}}

\title{Technological networks and the\\
spread of computer viruses}

\author{%
Justin Balthrop,$^{1}$ Stephanie Forrest,$^{1,2\ast}$ M. E. J. Newman,$^{3}$\\
and Matthew M. Williamson$^{4}$\\
\\
\normalsize{$^1$Dept.\ of Computer Science, University of New Mexico,
Albuquerque, NM 87131, USA}\\
\normalsize{$^2$Santa Fe Institute, Santa Fe, NM 87505, USA} \\
\normalsize{$^3$Dept.\ of Physics and Center for the Study of Complex
Systems,}\\
\normalsize{University of Michigan, Ann Arbor, MI 48109, USA}\\
\normalsize{$^4$HP Laboratories Bristol, Filton Road,
Stoke Gifford,  Bristol, BS34 8QZ, UK}\\
\\
\normalsize{$^\ast$To whom correspondence should be addressed;
E-mail: forrest@cs.unm.edu.}
}

\date{}

\begin{document} 


\baselineskip24pt


\maketitle 


\begin{sciabstract}
  Computer infections such as viruses and worms spread over networks of
  contacts between computers, with different types of networks being
  exploited by different types of infections.  Here we analyze the
  structures of several of these networks, exploring their implications for
  modes of spread and the control of infection.  We argue that vaccination
  strategies that focus on a limited number of network nodes, whether
  targeted or randomly chosen, are in many cases unlikely to be effective.
  An alternative dynamic mechanism for the control of contagion, called
  throttling, is introduced and argued to be effective under a range of
  conditions.
\end{sciabstract}

\clearpage

Computer viruses and worms are an increasing problem for users of computers
throughout the world.  By some estimates 2003 was the worst year yet:
viruses halted or hindered operations at numerous businesses and other
organizations, disrupted ATMs, delayed airline flights, and even affected
emergency call centers. The Sobig virus alone is said to have caused more
than US\$30 billion in damage.\footnote{Citations documenting these events
are listed on \url{http://www.cs.unm.edu/~judd/virus.html}, as well as
citations to each of the viruses and worms mentioned in this paper.}  And
most experts agree that the damage could easily have been much worse.  For
example, Staniford~\etal\ describe a worm that could infect the entire
Internet in around 30 seconds~\cite{justin5}.  A worm of this scale and
speed could launch a distributed denial-of-service attack that would bring
the entire network to a halt, crippling critical infrastructure.

In this paper we use the term \defn{virus} to refer to malicious software
that requires help from computer users to spread to other computers.  Email
viruses, for instance, require someone to read an email message or open an
attached file in order to spread.  We use the term \defn{worm} for
infections that spread without user intervention.  Because they spread
unaided, worms can often spread much faster than viruses.

Computer infections such as viruses and worms spread over networks of
contacts between computers, with different types of networks being
exploited by different types of infections.  The structure of contact
networks affects the rate and extent of spreading of computer infections,
just as it does for human diseases~\cite{KW91,PV01a,LM01a,EMB02,NFB02}, and
understanding this structure is thus a key element in the control of
infection.

Both traditional and network-based epidemiological models have been applied
to computer contagion~\cite{KW91,PV01a,LM01a}.  Recent work has emphasized
the effects of a network's \defn{degree distribution}.  A network consists
of nodes or \defn{vertices} connected by lines or \defn{edges}, and the
number of edges connected to a vertex is called its \defn{degree}.  Of
particular interest are \defn{scale-free networks}, in which the degree
distribution follows a power-law, the fraction $p_k$ of vertices with
degree~$k$ falling off with increasing~$k$ as $k^{-\alpha}$ for some
constant~$\alpha$.  This structure has been reported for several
technological networks including the Internet~\cite{FFF99} and the world
wide web~\cite{AJB99,Kleinberg99b}.

Infections spreading over scale-free networks are highly resilient to
control strategies based on randomly vaccinating or otherwise disabling
vertices.  This is bad news for traditional computer virus prevention
efforts, which use roughly this strategy.  On the other hand, targeted
vaccination, in which one immunizes the highest-degree vertices, can be
very effective~\cite{AJB00,CNSW00}.

It is important to appreciate that these results rely crucially on the
assumption of a power-law degree distribution, and their derivation also
assumes that the contact patterns between nodes are static.  Many
technological networks relevant to the spread of viruses, however, are not
scale-free.  Vaccination strategies focusing on highly connected network
nodes are unlikely to be effective in such cases.  Furthermore, network
topology is not necessarily constant.  In many cases the topology depends
on the replication mechanism employed by a virus and can be manipulated by
virus writers to circumvent particular control strategies that we attempt.
If, for instance, targeted vaccination strategies were found to be
effective against viruses spreading over scale-free networks, the viruses
might well be rewritten so as to change the structure of the network to
some non-scale-free form instead.

To make these ideas more concrete, we consider here four illustrative
technological networks, each of which is vulnerable to attack: (A)~the
network of possible connections between computers using the Internet
Protocol~(IP); (B)~a network of shared administrator accounts for desktop
computers; (C)~a network of email address books; (D)~a network of email
messages passed between users.
  
In network~A, the IP network, each computer has a 32-bit IP address and
there is a routing infrastructure that supports communication between any
two addresses.  We consider the network in which the nodes are IP addresses
and two nodes are connected if communication is possible between the
corresponding computers.  Although some segments of the IP address space
are invisible to others, the portion of the network within, for instance, a
corporation will generally be almost completely connected.  Many epidemics
spread over this IP network.  Notable examples include the Nimda and
SQLSlammer worms.

Network~B is a product of the common operating system feature that allows
computer system administrators to read and write data on the disks of
networked machines.  Some worms, including Nimda and Bugbear, can spread by
copying themselves from disk to disk over this network.

Network C is a directed graph with nodes representing users and a
connection from user~$i$ to user~$j$ if $j$'s email address appears in
$i$'s address book.  Many email viruses use address books to spread
(e.g.,~ILoveYou).  A closely related network is network~D, which is an
undirected graph in which the nodes represent computer users and two users
are connected if they have recently exchanged email.  Viruses such as Klez
spread over this network.

\begin{figure}[t]
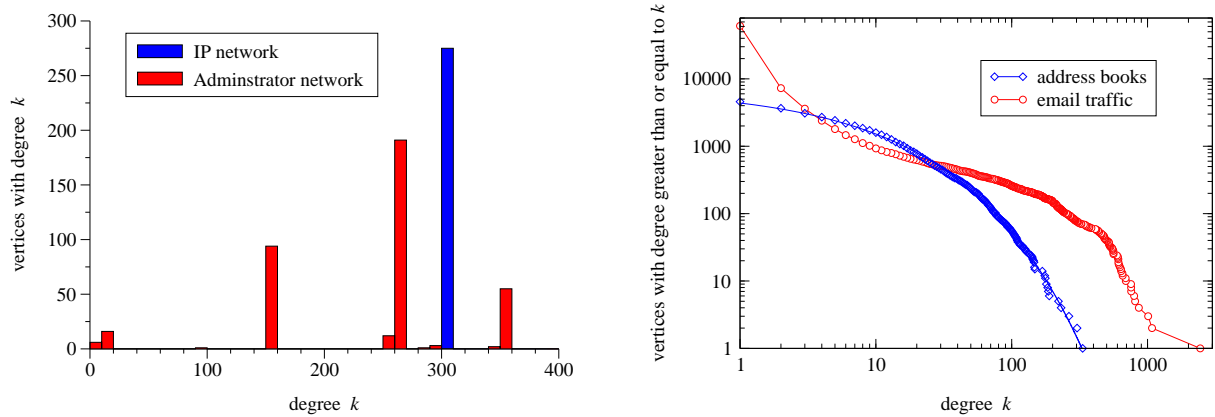

\begin{flushleft}
\resizebox{7.5cm}{!}{\includegraphics{ipadmin.eps}}\hfill
\resizebox{7.5cm}{!}{\includegraphics{email.eps}}
\end{flushleft}
\caption{Left: degree distributions for the IP and administrator networks.
The administrator network data were collected on a system of 518 users of
382 machines at a large corporation.  Computers in this network are
connected if any user has administrative privileges on both computers.
Right: Cumulative degree distributions for the email address book and email
traffic networks, plotted on logarithmic scales.  The email address book
data were collected from a large university~\cite{NFB02}.  The email
traffic data were collected for complete email activity of a large
corporate department over a four-month period.}
\label{fig:combined}
\end{figure}

Fig.~\ref{fig:combined} shows measured degree distributions for examples of
each of the four networks.  In network~A all vertices have the same degree,
so the distribution has a single peak at this value (blue histogram).  In
network~B, the distribution consists of four discrete peaks, presumably
corresponding to different classes of computers, administrators, or
administration strategies (red histogram).

The two email networks have more continuous distributions and are shown as
cumulative histograms.  Although neither network has a power-law degree
distribution, both have moderately long tails, which suggests that targeted
vaccination strategies might be effective.  Calculations show however that
for the address-book network about 10\% of the highest-degree nodes would
need to be vaccinated to prevent an epidemic from spreading~\cite{NFB02}
while the email traffic network would require about~87\%.  The first of
these figures is probably too high for an effective targeted vaccination
strategy, and the second is clearly far too high.  (Targeted vaccination
would be entirely ineffective in the other two networks as well, because
the nodes are much more highly connected.)

The two email networks illustrate the way in which different virus
replication strategies can lead to different network topologies.  An email
virus could look for addresses in address books, thereby spreading over a
network with a topology like that of network~C, or it could look elsewhere
on the machine, giving a topology more like~D.  Another example is provided
by the Nimda virus, which infects web servers by targeting random IP
addresses, producing a network like network~A.  However, if the virus had a
more intelligent way of selecting IP addresses to attack (say, by
inspecting hyperlinks), then it might spread over a topology more like that
of the world wide web, which is believed to have a power-law degree
distribution~\cite{AJB99,Kleinberg99b}.

A control strategy is needed that is immune to changes in network topology
and that does not require us to know the mechanisms of infections before an
outbreak.  Control strategies need to be highly effective against malicious
infection but largely transparent to legitimate activities.  A number of
methods for containing computer epidemics have been proposed
\cite{MooreEtal03a}.  One such strategy is \emph{throttling}, first
introduced for the control of misbehaving
programs~\cite{Somayaji-00-USENIX} and recently extended to computer
network connections~\cite{Williamson-03-C}.  In this context, throttling
limits the number of new connections a computer can make to other machines
in a given time period.  Because it works by limiting spreading rates
rather than stopping spread altogether, the method does not completely
eliminate infections but only slows them down.  Frequently however this is
all that is necessary to render a virus harmless or easily controlled by
other means.

The network traffic generated by today's viruses and worms is fundamentally
different from normal network traffic.  For a virus to spread it needs to
propagate itself to many different machines, and to spread quickly it must
do so at a high rate.  For example, the Nimda worm attempts to infect web
servers at a rate of around 400 new machines per second, which greatly
exceeds the normal rate of connections to new web servers of about one per
second or slower~\cite{Williamson-02-ACSAC}.  A throttling mechanism that
limited connections to new web servers to about one per second would slow
Nimda by a factor of 400 without affecting typical legitimate traffic.
This could easily be enough to change a serious infection into a minor
annoyance, which could then be eliminated by traditional means.  Slowing
the spread of Nimba by a factor of 400 (from a day to over a year) would
have allowed plenty of time to develop and deploy signatures and
prophylactic software patches.  (Of course, if throttling were implemented
on only a subset of the nodes in a network, then infections could spread
more easily.)  In addition to reducing virus spread, throttling has the
practical benefit of reducing the amount of traffic generated by an
epidemic, thus reducing demand on networking equipment---often the primary
symptom of an attack.

To summarize, approaches for controlling epidemics in computer networks are
unlikely to be generally effective if they rely on detailed assumptions
about network topology.  Moreover, the topology of contact networks is
often dependent on a virus's replication and infection strategy.  Better
and more general approaches to restricting contagion are needed.
Throttling provides one example of how rapidly spreading infections may be
controlled in the future.  Throttling can be effective for all network
topologies, although only if normal network traffic is significantly
different from that generated by spreading epidemics.  Throttling does not
eliminate computer epidemics; it simply causes them to spread more slowly,
allowing time for the slower mechanisms of conventional prevention and
clean-up.

The disparity between the speed of computer attacks (machine and network
speed) and the speed of manual response (human speed) has increased in
recent years.  As this trend continues, automated mechanisms like
throttling will likely become an essential tool for controlling those
attacks, complementing the largely manual approach of software patching in
use today.  The idea of rate limits is not specific to viruses, and could
be applied to any situation in which an attack or cascading failure occurs
faster than possible human response.

\vspace{1cm}
{\small The authors gratefully acknowledge Jeff Gassaway for help
collecting the email address book data set, Chaz Hickman for the
administrator data set, and Joshua Tyler and Bernardo Huberman for the
email traffic data set.  This work was funded in part by the James S.
McDonnell Foundation, the National Science Foundation, the Defense Advanced
Projects Research Agency, Intel Corp., and the Santa Fe Institute.}

\end{document}